\documentclass[preprint, prX]{revtex4}
\usepackage{graphicx}
\usepackage{dcolumn}
\usepackage{bm}
\usepackage{latexsym}
\usepackage{mathrsfs}
\usepackage{amsmath}
\usepackage{subfigure}
\usepackage[colorlinks]{hyperref}

\begin{document}

\title{The bouncing cosmology with $F(R)$ gravity and its reconstructing}

\author{Ali R. Amani}
\email{a.r.amani@iauamol.ac.ir }
\affiliation{\centerline{Department of Physics, Faculty of Sciences, Ayatollah Amoli Branch, Islamic Azad University,}\\ P.O. Box 678, Amol, Iran.}

\date{\today}

\pacs{98.80.-k; 95.36.+x; 04.50.Kd}
\keywords{Dark energy, $F(R)$ gravity; Equation of State parameter; Dark energy.}


\begin{abstract}
In this paper, we study $F(R)$ gravity by Hu--Sawicki model in Friedmann--Lema\^{\i}tre--Robertson--Walker (FLRW) background. The Friedmann equations are calculated by modified gravity action, and then the obtained Friedmann equations are written in terms of standard Friedmann equations. Next behavior of bouncing cosmology is investigated in the modified gravity model, i.e., this behavior can solve problem of non-singular in standard Big Bang cosmology. We plot the cosmological parameters in terms of cosmic time and then bouncing condition is investigated. In what follows, we reconstruct the modified gravity by redshift parameter, and also graphs of cosmological parameters are plotted in terms of redshift, in which the figures show us an accelerated expansion of Universe. Finally, the stability of the scenario is investigated by a function as sound speed, and the graph of sound speed versus redshift show us that there is the stability in late time.
\end{abstract}
\maketitle


\section{Introduction}\label{I}

Since the general relativity can good explain on many specifications of Universe by observational evidence and various theories. Then, we should concentrate our attention to other unknown issues. One of the unknown issues is non-singularity in the standard big bang cosmology. To solve this issue we need to introduce a scientific model for the known Universe, so that it describes the Universe as oscillatory. This means that our Universe is as a result of the collapse of a previous universe \cite{Ashtekar-2006}. On the other hand, a new idea was proposed named bouncing Universe that is able to solve non-singularity of the big bang cosmology \cite{Peter-2002, Brandenberger-2002}. Also, the bouncing Universe was studied in brane cosmology \cite{Kanti-2003} and in vector field \cite{Sadeghi-2010}. Interpretation of the bouncing Universe is that when the bounce occurs the Universe is entered to big bang era, and the Universe transfers from an initial contracting phase to an expanding phase so that at this point the Hubble parameter transits from $H < 0$ to $H > 0$ and we have in bounce point $H = 0$ \cite{Carloni-2006, Novello-2008, Sadeghi-2009, Cai-2012}.

Nowadays, many articles have been written due to observational evidence and various theories about expansion of the Universe which the expanding is undergoing an accelerating phase. The discovery of this topic was first arisen in type Ia supernova \cite{Riess-1998}, associated with large scale structure  \cite{Tegmark-2004} and cosmic microwave background  \cite{Bennet-2003}. We note that the accelerated expansion is due to a mysterious energy so-called dark energy which is more than $70 \%$ of the total energy in our Universe. The Universe dominates with a perfect fluid by a negative pressure and an equation of state (EoS) parameter which is less than $-1$ so-called phantom phase. There are many candidates to describe dark energy scenario so that we can introduce some of them such as the cosmological constant \cite{Weinberg-1989}, the scalar fields (including quintessence, phantom, quintom, tachyon and etc) \cite{Kamenshchik-2001, Caldwell-2002, Amani-2011, Sadeghi1-2009, Setare-2009, Setare1-2009}, holographic models \cite{Wei-2009, Amani1-2011, Amani-2015}, modified models \cite{Amani1-2015}, interacting models \cite{Amani-2013, Amani-2014, Naji-2014} and braneworld models \cite{Sahni-2003, Setare-2008, Brito-2015}.

Among these models, modified gravity theory has advantages in comparison with other models, because it prevents the complicated computation of numerical solutions, and also it is consistent with recent observations for late accelerating Universe and dark energy.
One of the modified gravity models is accomplished by changing the Ricci scalar $R$ to $F(R)$ with an arbitrary function in the gravitational action which this model is named $F(R)$ gravity theory \cite{Nojiri-2003}. The $F(R)$ gravity is a good alternative instead of the standard gravity model as a source of dark energy. As it just mentioned, the Einstein--Hilbert action is written in terms of two terms $F(R)$ and matter Lagrangian, and will obtain Friedmann equations in Friedmann--Lema\^{\i}tre--Robertson--Walker (FLRW) metric.

In this paper, we will avoid the initial singularity in the big bang theory by $F(R)$ theory and bouncing model. For this purpose, we consider the bouncing cosmology with $F(R)$ gravity by Hu--Sawicki model \cite{Hu-2007}, and we will show that there is an accelerated phase shift from an initial contracting phase to an expanding phase. The issue is demonstrates by scale factor $a(t)$, which the scale factor derivative ($\dot{a} <0$) decrease during the contracting phase and it ($\dot{a} >0$) increase in the expanding phase and also it ($\dot{a} =0$) is equal zero in the bounce point. The corresponding figures will confirm the aforesaid story.

 In what follows, we try to describe the $F(R)$ gravity as a source of dark energy to reconstruct by redshift parameter. Therefore, we will investigate a parametrization for $F(R)$ gravity with this motivation that we can describe the accelerated expansion of the Universe. Finally, we will investigate the stability of the model so that University is considered as a thermodynamic system in an adiabatic perturbation. Therefore, by using of an useful function named sound speed will study the stability of the model.

The paper is organized as follows:

In Sec. \ref{II}, we review $F(R)$ gravity model and obtain the Friedmann equations by using the corresponding action in FLRW metric. In Sec. \ref{III}, we investigate the bouncing behavior by Hubble parameter and the scale factor, and we obtain bouncing condition by HU--Sawicki model. In Sec. \ref{IV}, we reconstruct the current model with redshift parameter and then we will use a parametrization for precise description of the dark energy. Thereinafter, effective energy density and effective pressure of Universe will be written in terms of redshift, and will plot the cosmological parameters. In Sec. \ref{V}, we study stability of the model and investigate it in late time. Finally, in Sec. \ref{VI}, we present a short summary for this job.


\section{The $F(R)$ gravity theory}\label{II}

Let us start by considering coupling influence $F(R)$ gravity with matter in $4$-dimensional action as
\begin{equation}\label{e1}
S=\int d^4x \sqrt{-g}\left(F(R)+2 \kappa^2 \mathcal{L}_m\right),
\end{equation}
where $\kappa^2 = 8 \pi G$, $F(R)$ is an arbitrary function of Ricci scalar R, and $\mathcal{L}_m$ is a matter Lagrangian. By taking the variation of the action \eqref{e1} with respect to metric $g_{\mu \nu}$, one obtains
\begin{equation}\label{e2}
F_R(R) R_{\mu \nu}-\frac{1}{2} F(R) g_{\mu\nu}-\left(\nabla_\mu \nabla_\nu- g_{\mu\nu} \nabla^2\right) F_R(R) = \kappa^2 T_{\mu \nu}^{(m)},
\end{equation}
where $F_R = \frac{d F}{d R}$. We consider a spatially flat FLRW background as follows:
\begin{equation}\label{e3}
ds^2 = -dt^2 + a(t)^2\left(dr^2+r^2 d\Omega^2\right),
\end{equation}
where $a(t)$ is the scale factor. The Ricci scalar R is found as
\begin{equation}\label{e4}
R = 6 (2 H^2 + \dot{H}),
\end{equation}
where $H = \frac{\dot{a}}{a}$ is the Hubble parameter and a dot denotes a derivative with respect to cosmic time.\\
The two Friedmann equations obtain from \eqref{e2} as
\begin{subequations}\label{e5}
\begin{eqnarray}
3H^2 F_R &=& \kappa^2 \rho_m+\frac{1}{2}(R F_R-F)-3 H \dot{R} F_{RR},\label{e5-1}\\
-(3 H^2+2 \dot{H})F_R &=& \kappa^2 p_m+\frac{1}{2}(F-R F_R)+\dot{R}^2 F_{RRR}+2 H \dot{R} F_{RR}+\ddot{R} F_{RR},\label{e5-2}
\end{eqnarray}
\end{subequations}
where $\rho_m$ and $p_m$ are energy density and pressure of an Universe dominated by matter, respectively. By using the continuity equation $\nabla_\mu T^{(m){\mu \nu}} = 0$, and for a perfect fluid with the EoS parameter $p_m=\omega_m \rho_m$ finds
\begin{equation}\label{e6}
\dot{\rho}_m + 3 H (1 + \omega_m)\rho_m = 0,
 \end{equation}
where solution of the differential equation becomes
\begin{equation}\label{e7}
\rho_m = \rho_{m_0} a^{-3(1+\omega_m)}.
\end{equation}

Now by comparing the current model with the standard Friedmann equations ( $\rho_{eff} =  \frac{3}{\kappa^2} H^2$ and $p_{eff} = -\frac{1}{\kappa^2} (3 H^2+2 \dot{H})$ ), we can rewrite Eqs. \eqref{e5} in terms of the effective energy density and effective pressure as
\begin{subequations}\label{e8}
\begin{eqnarray}
\rho_{eff} = \rho_{F(R)} +\rho_m = \frac{1}{\kappa^2} \left(\kappa^2 \rho_m+3 H^2 (1-F_R)+\frac{1}{2}(R F_R-F)-3 H \dot{R} F_{RR}\right),\label{e8-1}\\
\begin{aligned}
p_{eff} = p_{F(R)} +p_m = \frac{1}{\kappa^2} \bigg(\kappa^2 p_m+(3 H^2+2 \dot{H})(F_R-1)+&\frac{1}{2}(F-R F_R)+\dot{R}^2 F_{RRR}\label{e8-2}\\
&+2 H \dot{R} F_{RR}+\ddot{R} F_{RR}\bigg),
\end{aligned}
\end{eqnarray}
\end{subequations}
where
\begin{subequations}\label{e9}
\begin{eqnarray}
&\rho_{F(R)} = \frac{1}{\kappa^2} \left(3 H^2 (1-F_R)+\frac{1}{2}(R F_R-F)-3 H \dot{R} F_{RR}\right),\label{e9-1}\\
&p_{F(R)} = \frac{1}{\kappa^2} \left((3 H^2+2 \dot{H})(F_R-1)+\frac{1}{2}(F-R F_R)+\dot{R}^2 F_{RRR}+2 H \dot{R} F_{RR}+\ddot{R} F_{RR}\right).\label{e9-2}
\end{eqnarray}
\end{subequations}

The conservation equation can be obtained by using Eqs. \eqref{e8}  as
\begin{equation}\label{e10}
\dot{\rho}_{eff} + 3 H (1 + \omega_{eff})\rho_{eff} = 0,
 \end{equation}
where
 \begin{equation}\label{e10-1}
\omega_{eff} = \frac{p_{ef f}}{\rho_{ef f}} = -1-\frac{2 \dot{H}}{3 H^2},
\end{equation}
is the effective EoS parameter. We explorer bouncing behavior in next section.

\section{Bouncing behavior}\label{III}

In this section, we will describe the bouncing conditions in $F(R)$ gravity model. When a bouncing Universe occurs that Universe transfers of fluctuations from an initial contracting phase to an expanding phase. This phase shift give rise to a solution of non-singular in the standard Big Bang cosmology \cite{Carloni-2006, Novello-2008, Sadeghi-2009, Cai-2012}.

 Therefore, for a successful bounce, the Hubble parameter passes from $H < 0$ to $ H > 0$ and in bounce point $H = 0$. Also we can express issue of bouncing Universe in terms of the scale factor, i.e.,  during the contracting phase we have a decrease for the scale factor as $\dot{a} < 0$, and in the expanding phase we have a increase as $\dot{a} > 0$, and in the bounce point $\dot{a} =  0 $ and around this point $\ddot{a}  >  0$.

Since the Hubble parameter passes from $H<0$ to $H>0$ in terms of time evolution for the bouncing Universe, then the derivative of Hubble parameter must be bigger than zero in bounce point as
\begin{equation}\label{e11}
\dot{H}_{bounce}=-\frac{\kappa^2}{2}(1+\omega_{eff}) \rho_{eff} > 0,
\end{equation}
from this condition we deduce $\omega_{eff}<-1$ and $\rho_{eff}>0$.

Now we can obtain the bounce condition for the model in bounce point. In that case, we must have $\dot{H}_{bounce}>0$ and $H_{bounce}=0$, so that the bouncing condition  \eqref{e11} in the bounce point by Eqs. \eqref{e8} be deduced as
\begin{equation}\label{e12}
\dot{H}_{bounce}=\frac{\kappa^2(\rho_m+p_m)+\dot{R}^2 F_{RRR}+\ddot{R} F_{RR}}{-2 F_R}>0.
\end{equation}
In the following, we will study our model by a particular $F (R)$ model that one was first proposed by Hu and Sawicki \cite{Hu-2007}. The Hu-Sawicki model can create the late time accelerated Universe, and also corresponded with a good approximation by the $\Lambda C D M$ model in the large redshift. We can write the corresponding model by \cite{HWLK}
\begin{equation}\label{e18}
F (R) =R-\mu\,{\it R_c}\, \left( 1- \left( 1+{\frac {{R}^{
2}}{{{\it R_c}}^{2}}} \right) ^{-n} \right)
\end{equation}
\begin{figure}[h]
\begin{center}
\subfigure{\includegraphics[scale=.3]{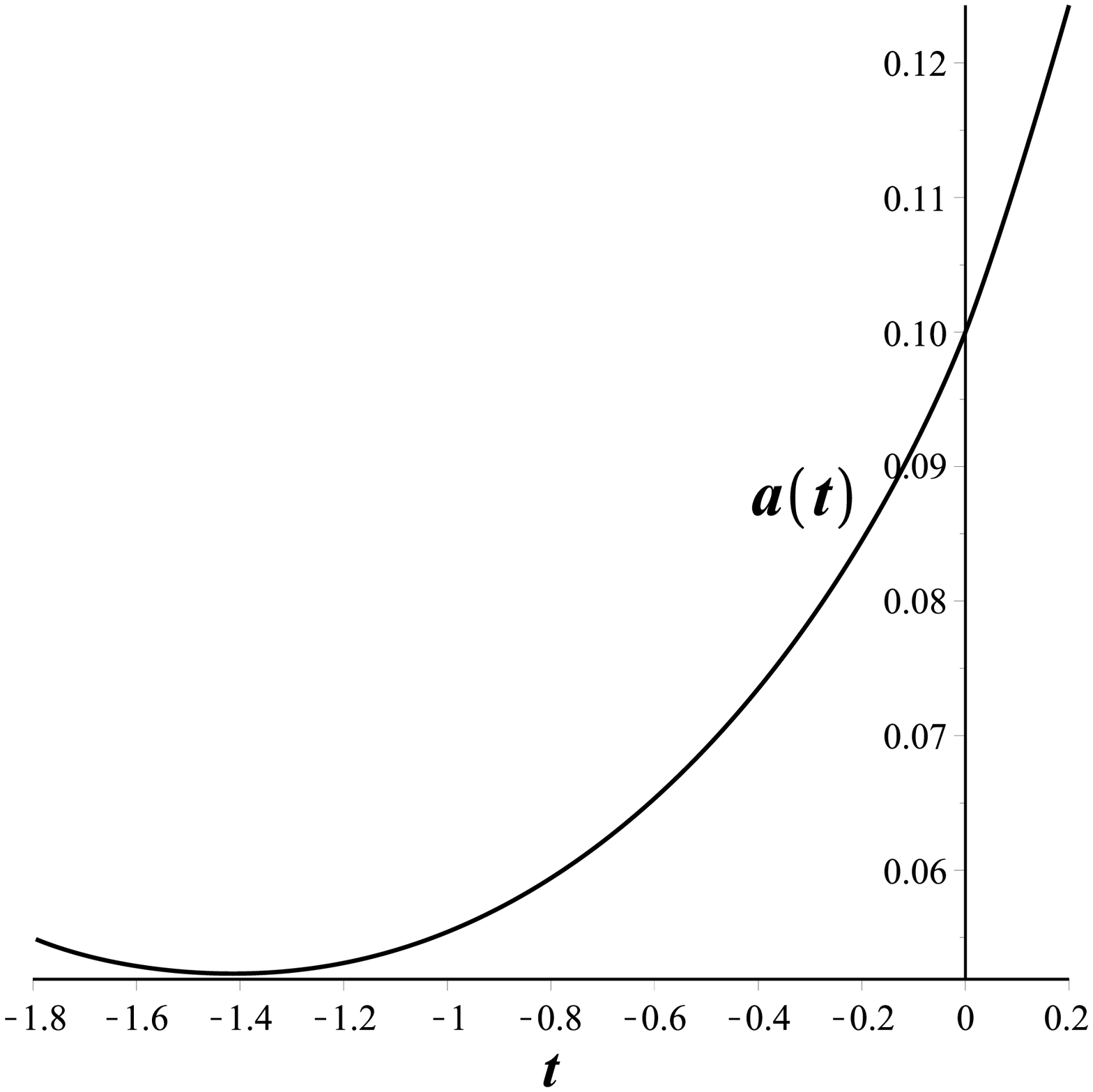}\label{fig1-1}}
\subfigure{\includegraphics[scale=.3]{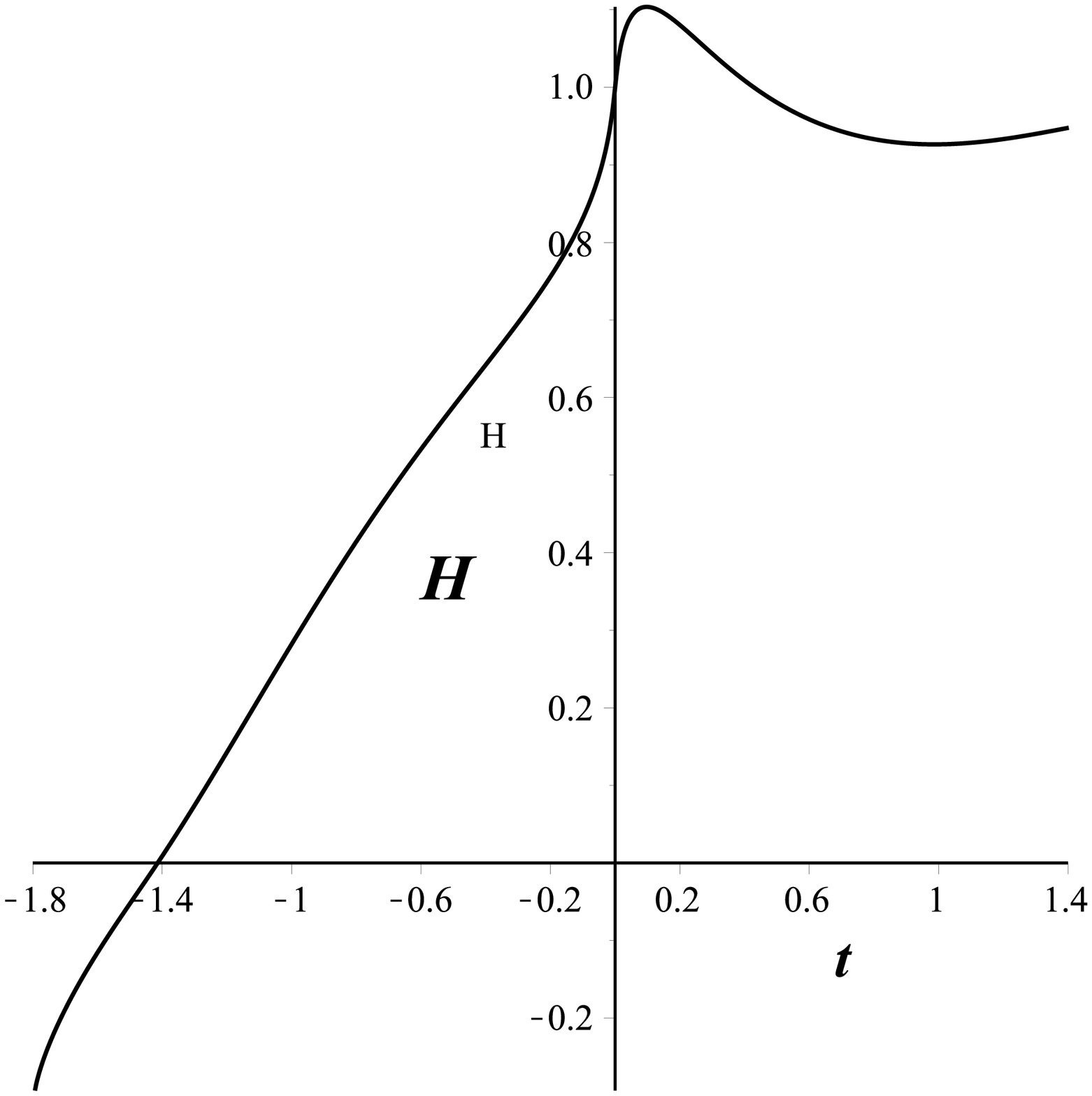}\label{fig1-2}}
\caption{Graph of the scale factor in terms of cosmic time by choosing $\kappa = 0.25$, $\rho_{m_0} = 1.5$, $\mu = 0.5$, $R_c = 5$, $\omega_m = -2.5$ and
$n = 2$ with $a(0) = 0.1$, $a'(0) = 0.1$, $a''(0) = 0.5$ and $a'''(0) = 0.5$.}\label{fig1}
\end{center}
\end{figure}
\begin{figure}[h]
\begin{center}
\subfigure{\includegraphics[scale=.3]{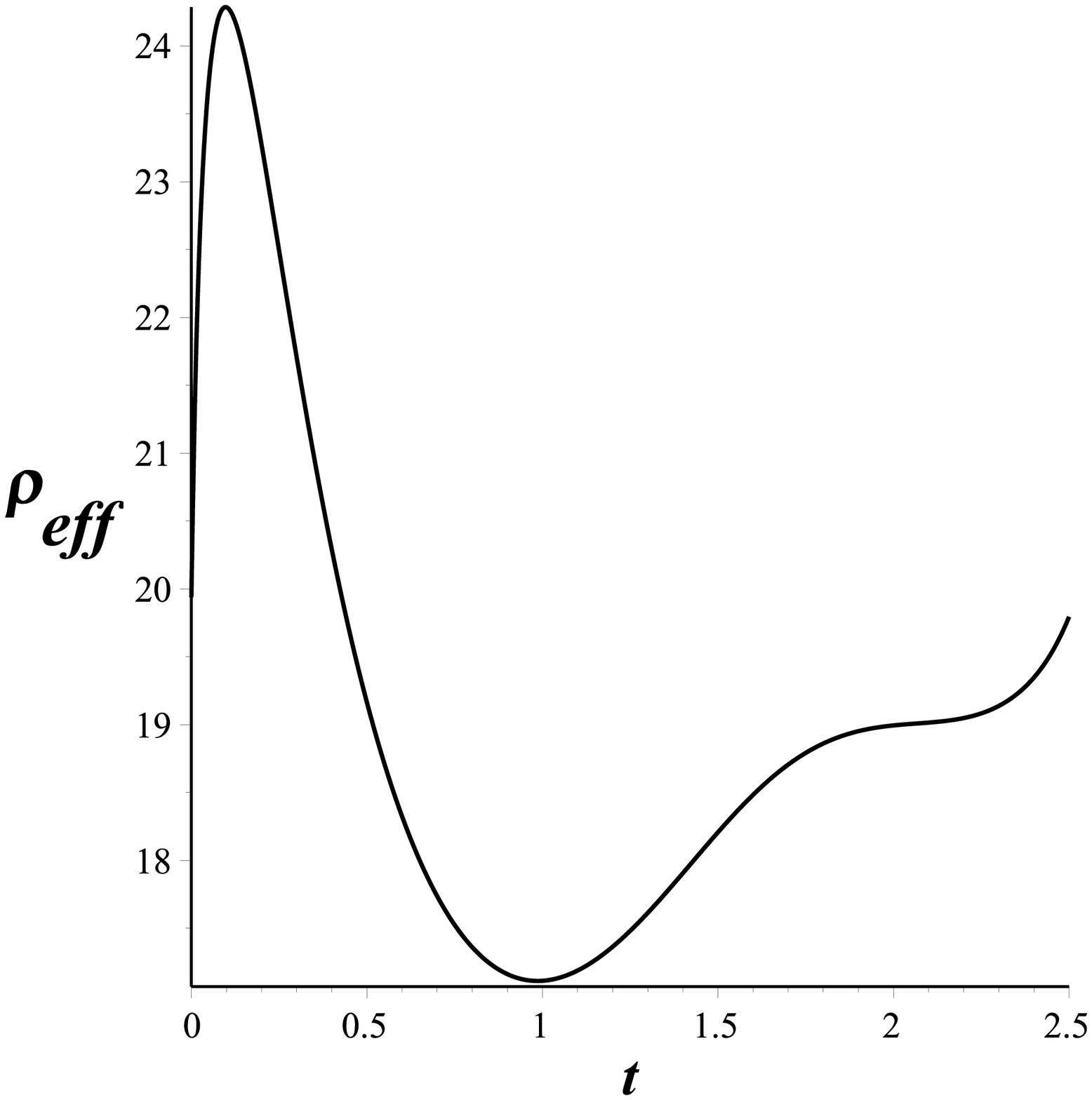}\label{fig2-1}}
\subfigure{\includegraphics[scale=.3]{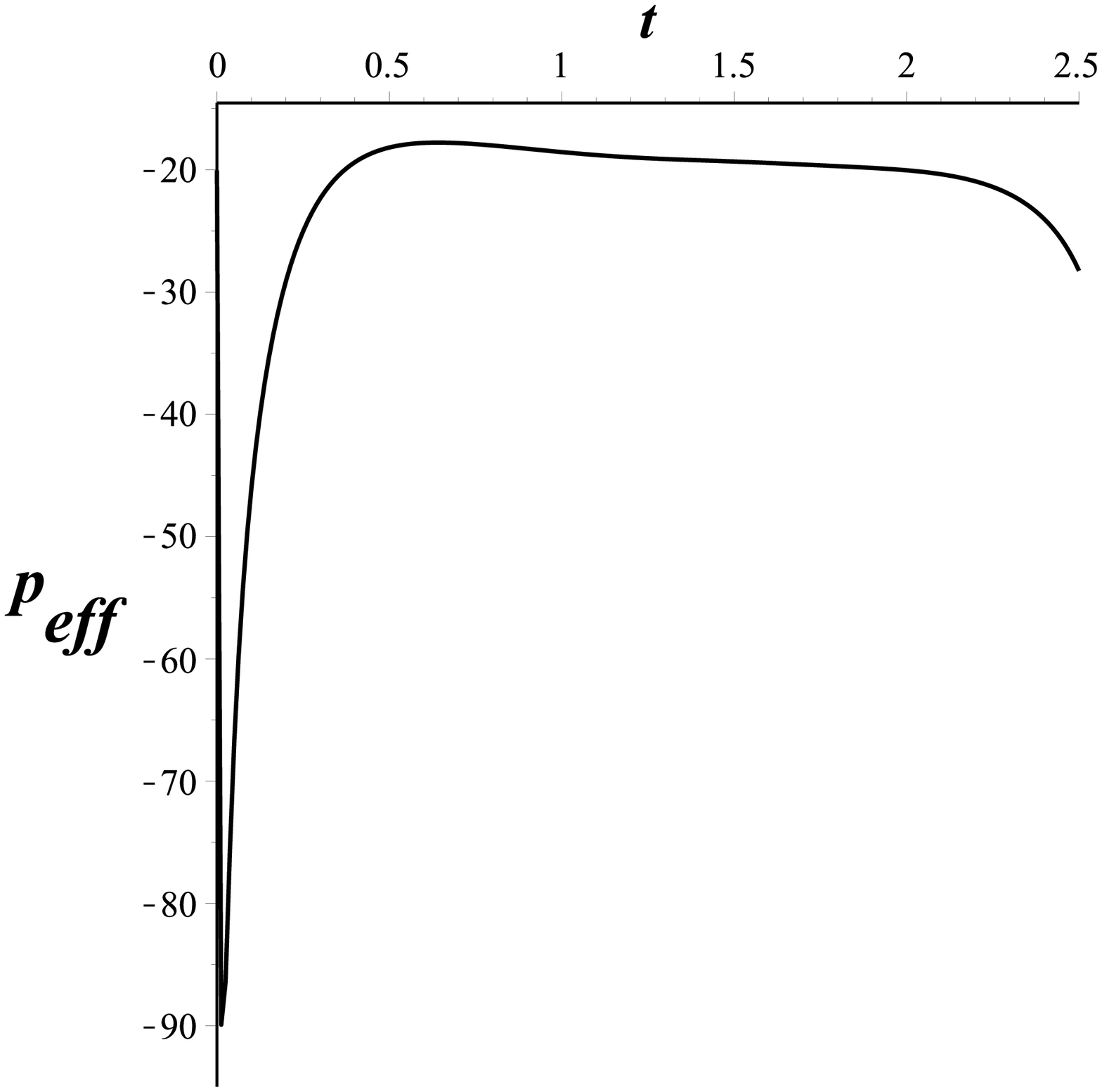}\label{fig2-2}}
\caption{Graph of the density energy and the pressure in terms of cosmic time by choosing $\kappa = 0.25$, $\rho_{m_0} = 1.5$, $\mu = 0.5$, $R_c = 5$, $\omega_m = -2.5$ and
$n = 2$ with $a(0) = 0.1$, $a'(0) = 0.1$, $a''(0) = 0.5$ and $a'''(0) = 0.5$.}\label{fig2}
\end{center}
\end{figure}
\begin{figure}[h]
\begin{center}
\includegraphics[scale=.3]{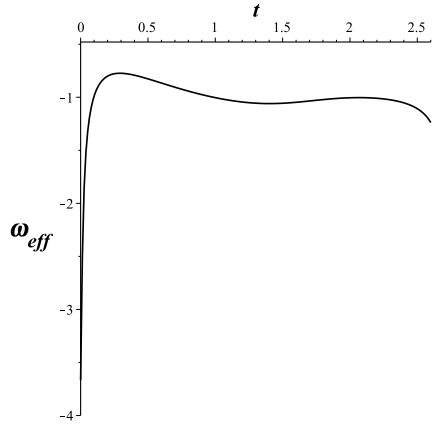}
\caption{Graph of the EoS parameter in terms of cosmic time by choosing $\kappa = 0.25$, $\rho_{m_0} = 1.5$, $\mu = 0.5$, $R_c = 5$, $\omega_m = -2.5$ and
$n = 2$ with $a(0) = 0.1$, $a'(0) = 0.1$, $a''(0) = 0.5$ and $a'''(0) = 0.5$.}\label{fig3}
\end{center}
\end{figure}
where $\mu$, $R_c$ and $n$ are free parameters, and use the natural units as $c = \hbar = m_p = 1$. By inserting \eqref{e18} into Eqs. \eqref{e8}, and numerical solution of the obtained Friedmann equations we can draw the cosmological parameters in terms of cosmic time in Figs. \ref{fig1}-\ref{fig3}. The motivation of this choice is based on crossing of EoS over phantom-divide-line, and finding bouncing condition. We note that the bouncing behavior can be observed in the Fig. \ref{fig1}, i.e., on the one hand, we can see that the Hubble parameter passes from the bounce point ($H(t \sim -1.4)=0$) as $H<0$ to $H>0$, and on the other hand, we see that the scale factor is minimum or $\dot{a}(t \sim -1.4)=0$.

The Fig. \ref{fig2} shows $\rho_{eff} > 0$ and $p_{eff} < 0$, which these show a accelerated Universe. Also we can observe variation of the EoS with respect to cosmic time in \ref{fig3}, and one show crossing over phantom-divide-line that the issue corresponds to observational data of late time \cite{Lazkoz-2005, SNES}. \\
In next section we will reconstruct the aforesaid model using the concept of redshift.


\section{Reconstructing by redshift parameter}\label{IV}

In this section, we are going to investigate the model by redshift parameter. One can be introduced in terms of the scale factor as $z=\frac{a_0}{a(t)}-1$ in which $a_0$ is the value of the scale factor at the present Universe, or other words redshift is zero in the late cosmic time. Therefore, dimensionless parameter $r(z)=\frac{H(z)^2}{H_0^2}$ is introduced so that $H_0=71\pm3\,km\,s^{-1}\,Mpc^{-1}$ is the Hubble parameter of current Universe. In order to rewrite the aforesaid Friedmann equations in terms of redshift, the differential form with respect to cosmic time becomes
\begin{equation}\label{e19}
\frac{d}{dt}=-H(1+z) \frac{d}{dz},
\end{equation}
then, Eqs. \eqref{e4} and \eqref{e8} are written in terms of $z$ in the following form
\begin{equation}\label{e19}
R=12H_0^2 r-3 H_0^2 (1+z) r',
\end{equation}
\begin{eqnarray}\label{rhototz}
\begin{aligned}
\rho_{eff}=\frac{1}{\kappa^2} \Big[\kappa^2 \rho_m+3 H_0^2 r(F_R(z)+1)- \frac{3}{2}H_0^2(1+z) r' F_R(z)+ 3 H_0^2(1+z) r &F'_R(z)\\
&-\frac{1}{2} F(z) \Big],
\end{aligned}
\end{eqnarray}
\begin{eqnarray}\label{ptotz}
\begin{aligned}
p_{eff}=\frac{1}{\kappa^2} \Big[\kappa^2 p_m+\frac{1}{2} & H_0^2 (1+z) r' (F_R(z)+2) - 3 H_0^2 r (F_R(z)+1) + \frac{1}{2}F(z)\\
&+H_0^2 (1+z)^2 r F''_R(z) + \frac{1}{2}H_0^2 (1+z)^2 r' F'_R(z) - H_0^2 (1+z) r F'_R(z) \Big],
\end{aligned}
\end{eqnarray}
where the prime denotes a derivative with respect to $z$. The matter energy density and functions $F(R)$ and its derivative with respect to $R$ are written by
\begin{eqnarray}
  & \rho_m =  \rho_{m_0} \left(\frac{a_0}{1+z}\right)^{-3(1+\omega_m)}, \\
  &F (z) =(12H_0^2 r-3 H_0^2 (1+z) r')-\mu\,{\it R_c}\, \left( 1- \Big( 1+{\frac {{(12H_0^2 r-3 H_0^2 (1+z) r')}^{
2}}{{{\it R_c}}^{2}}} \Big) ^{-n} \right), \\
  &F_R(z)=1-2 n \mu \frac{12H_0^2 r-3 H_0^2 (1+z) r'}{R_c}\left(1+\frac{(12H_0^2 r-3 H_0^2 (1+z) r')^2}{R_c^2}\right)^{-(n+1)}.
\end{eqnarray}

Now, in order to reconstruct the model as a origin of dark energy, we must introduce function $r(z)$ so that one is fitted with observational data of supernova \cite{AAST, DHMS}. For this purpose, one of the fitting functions is frequently considered as third-degree polynomial parametrization in terms of redshift, i.e., \cite{EJCO, UALA}
\begin{equation}\label{rz}
  r(z)=\Omega_{{m_{{0}}}} \left( 1+z \right) ^{3}+A_{{0}}+A_{{1}} \left( 1+z \right) +A_{{2}} \left( 1+z \right) ^{2},
\end{equation}
where $A_0=1 - A_1 - A_2 - \Omega_{m_0}$. It should be noted that the aforesaid parametrization matches with $\Lambda$CDM model when $A_1 = A_2 = 0$ and $A_0 =1 - \Omega_{m_0}$.  The best fit parameters are as $\Omega_{m_0} = 0.3$, $A_1=-4.16 \pm 2.53$ and $A_2 = 1.67 \pm 1.03$ \cite{Lazkoz-2005}. In this literature, we choose free parameter values of the parametrization as $\Omega_{m_0} = 0.3$, $A_0= 3.2$, $A_1=-3.5$ and $A_2 = 1$. We note that free parameters play a very important role in this job, so selected incentive is based on that effective energy density and effective pressure become positive and negative respectively, and also EoS crosses over the phantom divide line.

\begin{figure}[t]
\begin{center}
\includegraphics[scale=.27]{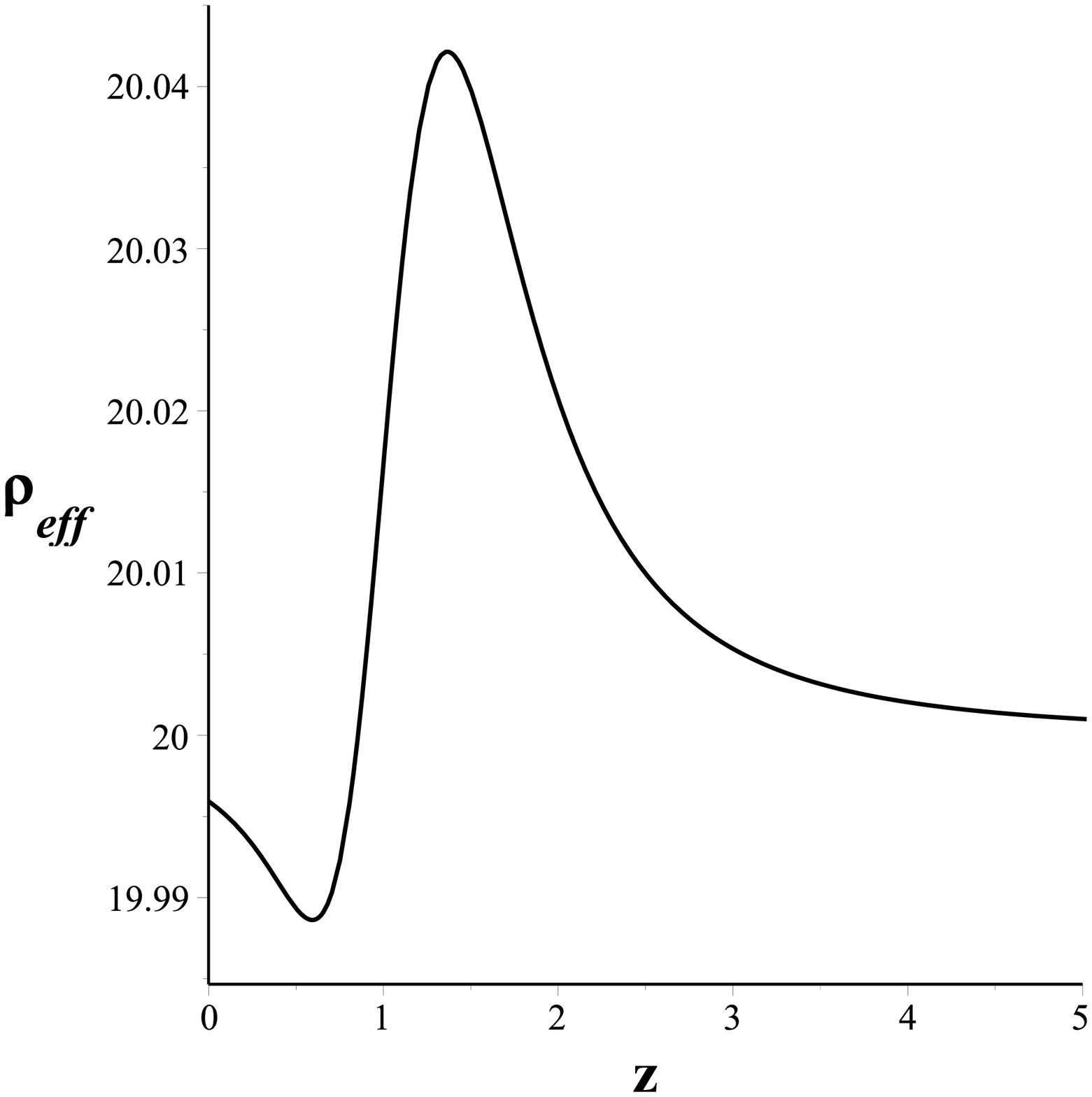}\includegraphics[scale=.27]{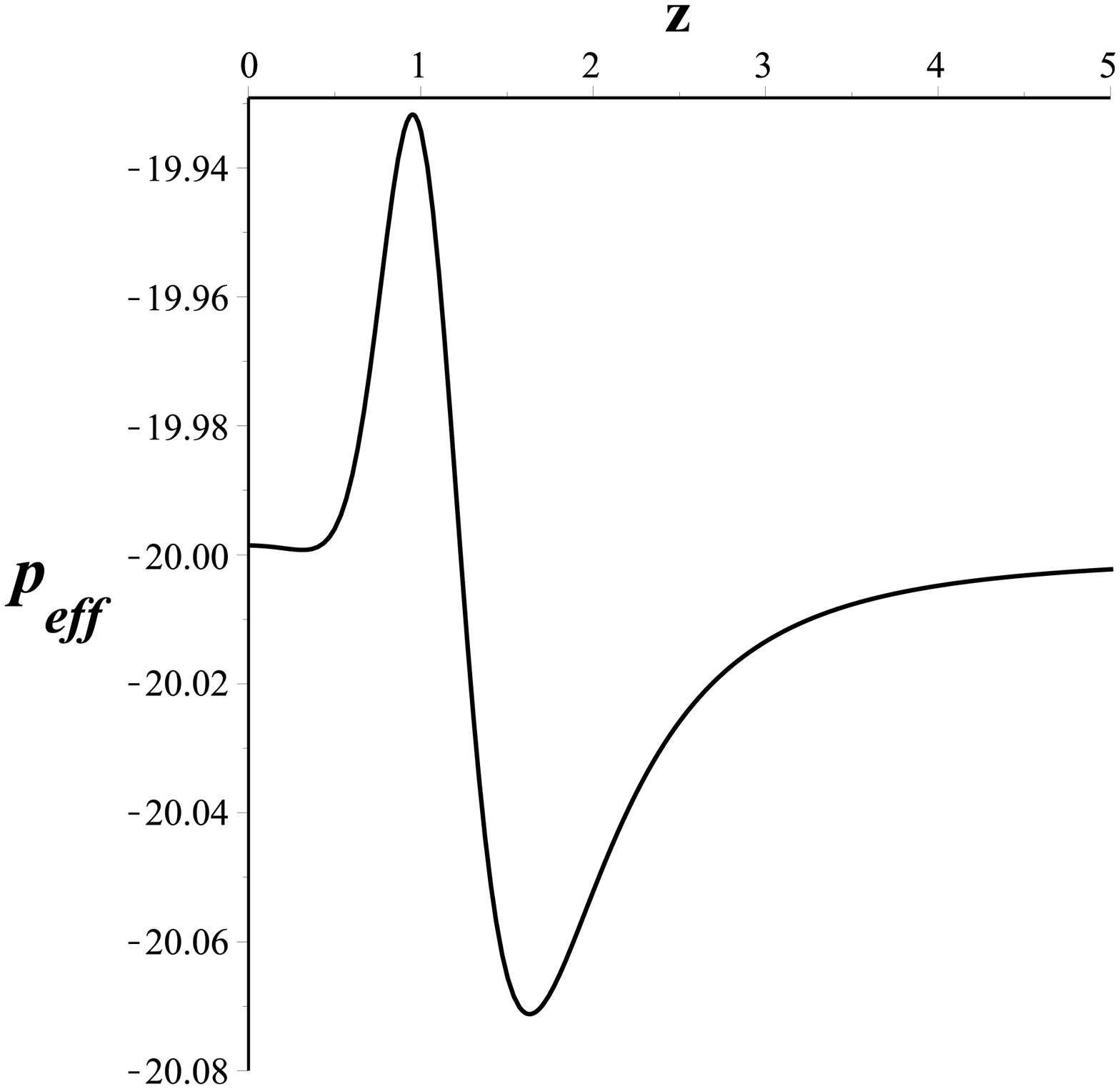}\includegraphics[scale=.27]{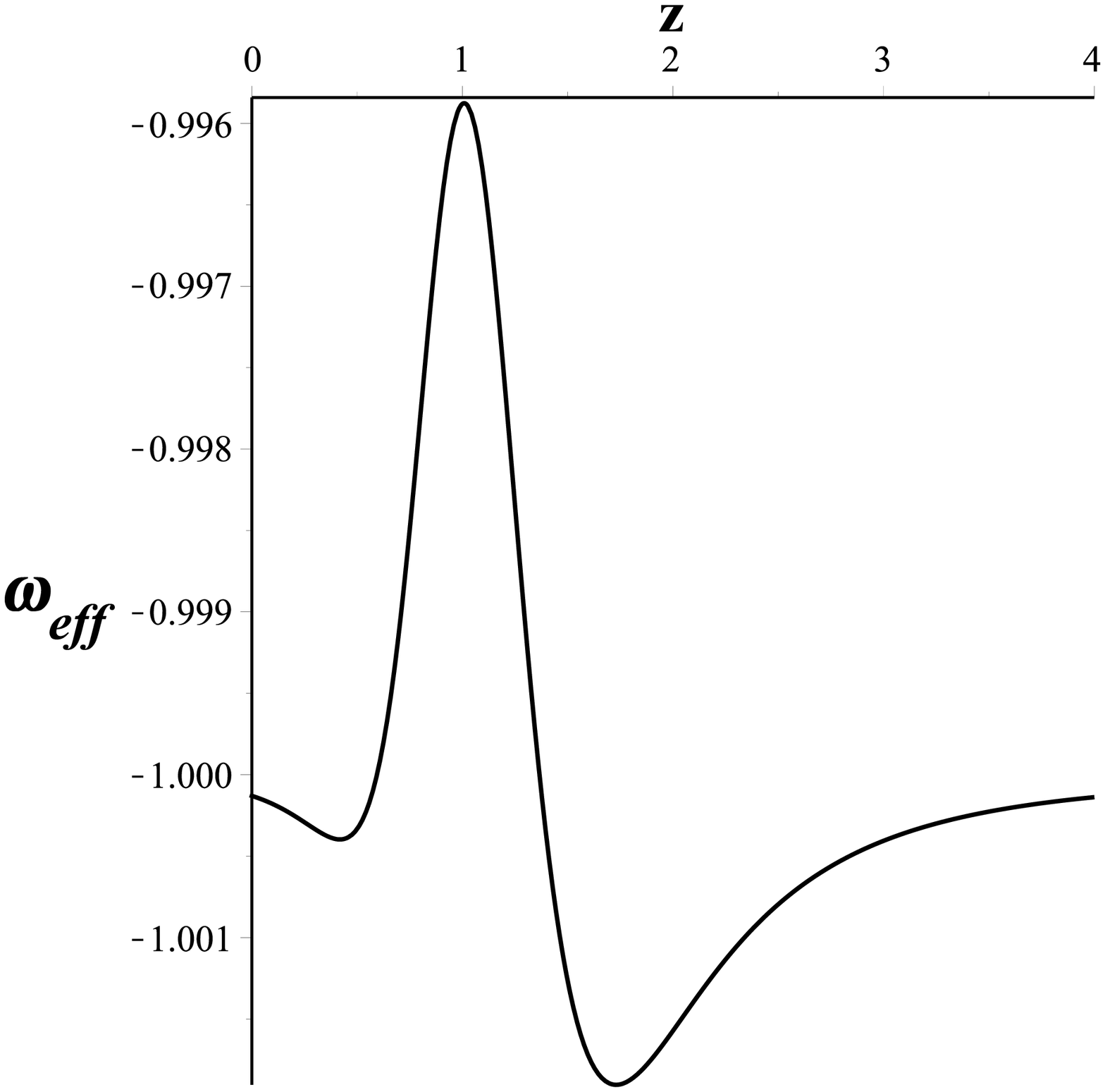}
\caption{Graphs of $\rho_{eff}$, $p_{eff}$ and $\omega_{eff}$ in terms of redshift by choosing $\kappa = 0.25$, $\rho_{m_0} = 1.5$, $\mu = 0.5$, $R_c = 5$, $\omega_m = -2.5$, $n = 0.5$ and $a_0 = 0.1$.}\label{fig4}
\end{center}
\end{figure}

In what follows, by substituting \eqref{rz} into Eqs. \eqref{rhototz} and \eqref{ptotz}, we can obtain cosmological parameters in terms of redshift. Therefore, we plot $\rho_{eff}$, $p_{eff}$ and $\omega_{eff}$ in terms of redshift accordance with Fig. \ref{fig4}. The variation of effective EoS shows us that the value of EoS is equal to about $-1.0001$ in late time ($z=0$), and this confirms condition $\omega_{eff}<-1$ as an accelerated expansion of the Universe. The obtained result verifies the result of Refs. \cite{Vasey-2007, Amanullah-2010} in a flat Universe.

\section{Stability}\label{V}
In this section, we will discuss on stability of the $F(R)$ gravity. Since Universe has been filled the perfect fluid, then we can consider it as a thermodynamic system. For this purpose, we will use the quantity of sound speed for the perfect fluid system. As we know the sound speed is introduced by an useful function as $c^2_s=\frac{d p_{eff}}{d \rho_{eff}}$, in which $p_{eff}$ and $\rho_{eff}$ are the effective energy density and effective pressure of Universe. Since the sound speed $c^2_s$ is positive in a thermodynamic system, hence the stability condition occurs when the function $c^2_s$ becomes bigger than zero. We note that a thermodynamic system can be described with adiabatic and non-adiabatic perturbations by quantities the effective energy density, effective pressure and entropy.

Now we consider the corresponding system as $p_{eff}=p_{eff}(S,\rho_{eff})$, and making perturbation with respect to the effective pressure we have
\begin{equation}\label{effpp}
  \delta p_{eff}=\left(\frac{\partial p_{eff}}{\partial S}\right)_{\rho_{eff}} \delta S+\left(\frac{\partial p_{eff}}{\partial \rho_{eff}}\right)_{S} \delta \rho_{eff} = \left(\frac{\partial p_{eff}}{\partial S}\right)_{\rho_{eff}} \delta S+ c^2_s \, \delta \rho_{eff},
\end{equation}
where the first term is related to a non-adiabatic process and second term is related to adiabatic process in the cosmological problem. Since we consider an adiabatic perturbation in cosmology, then variety of entropy becomes zero for the cosmological system, i.e., $\delta S = 0$. Therefore, we continue our research which is included with only adiabatic process.

Now in order to obtain function $c^2_s$, to differentiate the Eqs. \eqref{rhototz} and \eqref{ptotz} with respect to redshift, we obtain $c^2_s$ in terms of redshift. In that case, we plot the sound speed function in terms of redshift by numerical calculation as shown in Fig. \ref{fig4}. We can see in the corresponding figure the value of $c^2_s$ in late time ($z = 0$) with amount of approximately $0.07$ which satisfies condition $c^2_s>0$. Therefore, the Fig. \ref{fig4} shows us that there is a stability in late time, because value $c^2_s$ is positive for case $z=0$.
\begin{figure}[h]
\begin{center}
\includegraphics[scale=.3]{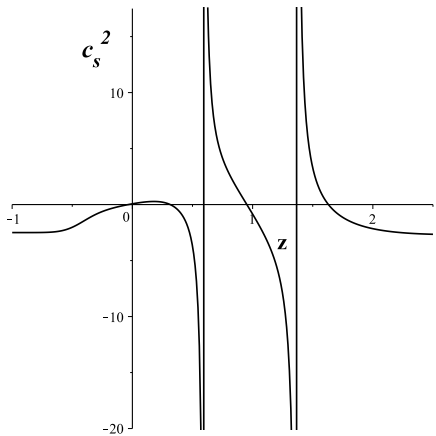}
\caption{Graph of $c_s^2$ in terms of redshift by choosing $\kappa = 0.25$, $\rho_{m_0} = 1.5$, $\mu = 0.5$, $R_c = 5$, $\omega_m = -2.5$, $n = 0.5$ and $a_0 = 0.1$.}\label{fig5}
\end{center}
\end{figure}

\section{Conclusions}\label{VI}

In this paper, we have studied $F(R)$ gravity by an arbitrary function of curvature called Hu--Sawicki model in FLRW metric. We wrote the corresponding action as the combination of the $F(R)$ gravity and the matter Lagrangian. The Friedmann equations have been obtained by the action and then we tried to write the obtained Friedmann equations as standard Friedmann equations. Next we considered the effective energy density and effective pressure of Universe so that one dominated by a perfect fluid. Then, we obtained the effective EoS by dividing these two functions $\rho_{eff}$ and $p_{eff}$ for the modified gravity. In what follows, we studied bouncing behavior for the scenario and obtained bouncing condition in bounce point and plotted the corresponding cosmological parameters in terms of cosmic time. Afterwards, we reconstructed the model by redshift parameter, and we used by the parametrization with function $r(z)$, and the cosmological parameters especially the Friedmann equations and the effective EoS have been written in terms of redshift $z$. Figures of effective energy density, effective pressure and effective EoS plotted in terms of redshift and they showed us that varieties of $\rho_{eff}$ and $p_{eff}$ are positive and negative respectively. On the other hand, variety of effective EoS showed us that the EoS crosses over the phantom phase, and late time amount of Eos is $-1.0001$. This result confirms accelerated expansion of Universe and also corresponds with observational data \cite{Vasey-2007, Amanullah-2010}. We noted that the free parameters play an very important role in the corresponding graphs, in which the motivation of these selections are based on positivity effective energy density and negativity effective pressure, and also crossing effective EoS over phantom divide line. Finally, in order to investigate the stability of the scenario, we tried to compute speed sound in terms of redshift, so we plotted the $c_s^2$ with respect to redshift and have obtained amount of sound speed in late time ($z = 0$) equal to $0.07$. Therefore, the Fig. \ref{fig4} showed us that there is the stability in late time, because function $c^2_s$ is bigger than zero in current time.


\section{Acknowledgements}
This work has been supported financially by Ayatollah Amoli Branch, Islamic Azad University, Amol, Mazandaran, Iran.



\end{document}